\documentstyle[aps,preprint,epsf,epsfig,rotate,prl]{revtex}
\newcommand{\be}{\begin{equation}}
\newcommand{\ee}{\end{equation}}
\newcommand{\ba}{\begin{eqnarray}}
\newcommand{\ea}{\end{eqnarray}}
\newcommand{\bdm}{\begin{displaymath}}
\newcommand{\edm}{\end{displaymath}}

\draft        
\begin{document}             
\title{A dilute mixture of atoms and molecules}
\author{S. K. Yip}
\address{Institute of Physics, Academia Sinica, Nankang, Taipei 11529, Taiwan}
\maketitle

\begin{abstract}
We study a dilute gas with two species of Fermionic atoms
of unequal concentrations, interacting via a short-range
interaction with one deeply bound state. 
We study the properties of this system under
the mean-field approximation.  We obtain
the effective interaction among the fermions
and bosons, and discuss the collective modes of
the system.


\end{abstract}

Recent experimental advances in trapping and cooling alkali atoms 
have already given us condensates of Bosons \cite{reviews}
and degenerate gases of Fermions \cite{DeMarco,Truscott01}.
Much theoretical and experimental efforts have now turned
to systems with trapped molecules,
with interesting observations and predictions on the equilibrium and
dynamical properties of this system.
\cite{Javanainen98,Drummond98,Timmermans99,Heinzen00,Wynar00,Calsamiglia01}
These molecules can have their constituent atoms 
be either Bosons or Fermions.\cite{Holland01}
The latter case, on which the rest of this paper
will concentrate, is also of particularly interest
when it is considered together with superfluidity 
arising from Cooper pairing of Fermions.
Indeed superfluid $^3$He and most known superconductors and 
have their superfluidity derived from Cooper pairs.
It is also widely believed that Fermionic pairing 
in diatomic molecules 
and in Cooper pairs are in fact two limits of a continuum.
\cite{Leggett80}
Much work has already been devoted to the cross-over between
these two limits. \cite{Melo93,Haussmann99} (and references therein)
There is also a lot of interest in producing pairs (within any part of
the above mentioned continuum) in
trapped atomic Fermions. \cite{Holland01}

We here would like to further our understanding of systems
consisting of "molecules".
We shall limit ourselves to "molecules" 
made up of two distinguishable Fermions,
interacting via a short range ($<< a$ below) interaction, 
in the s-wave state 
(Therefore by 'molecule' we simply mean the Boson made up by
a pair of Fermions).
These two partners of the molecule can in general be
atoms in different hyperfine spin projections or
internal states, or different atomic species.
For definiteness we shall restrict ourselves
to the first case and for simplicity
refer them hereafter as spin "up" and "down",
with concentrations $n_{\uparrow}$ and $n_{\downarrow}$.
 We shall study the properties of this
system when the two constituents of the molecule
are of different concentrations, say when 
$n_{\uparrow} > n_{\downarrow}$. 
Furthermore, we restrict ourselves to the case
where the size of the bound state $a$ is small,
such that $n_{\uparrow} a^3$ and $n_{\downarrow} a^3$ are 
both much smaller than $1$. Then at temperatures much
below the dissociation temperature $T_{\rm disso}$
all down spins are paired into molecules 
but there are left over (or excess) spin-up Fermions.
The Bosonic molecules and Fermionic atoms basically do not overlap and are
well-defined entities.
Thus one has a Bose-Fermi mixture where
 the Fermions are identical with one of the constituent
particles of the Bosons, with 
the Boson and Fermion concentrations $n_b = n_{\downarrow}$
and $\nu = n_{\uparrow} - n_{\downarrow}$ respectively.

We shall obtain the properties
of our Bose-Fermi mixture by applying many-body
techniques on the (more basic) constituent
spin-up and down Fermions
(generalizing ref \cite{Leggett80,Melo93,Haussmann99} 
-----  we shall also be following the notations of
the last reference quite closely, except that we
shall use the more common sign conventions for the single particle
propagators and self-energies).
This system is attractive theoretically in that all interactions
can be characterized by one interaction parameter, namely the
s-wave scattering length $a$ ($>0$) between the spin up and down Fermions,
or equivalently the binding energy $\epsilon_b$ of the molecule
($\epsilon_b = {\hbar^2 / m a^2}$, here $m$ is the mass of 
either of the Fermions).  
There are, however, several energy scales
and hence temperature regimes.  
As mentioned we shall limit ourselves to where the
temperature $T$ of the system is much lower than
the dissociation temperature  $T_{\rm disso}$ 
($\sim \epsilon_b$ up to some entropic factors).  
The relevant scales are thus $T$, 
the Bose-Einstein condensation temperature $T_{BEC}$ for
the Bosons and the Fermi temperature $T_F$ for the 
(left-over) Fermions ( $T_{BEC} \sim n_b^{2/3} \hbar^2 / m$ and
$T_F \sim \nu^{2/3} \hbar^2 / m$.
 Note they are much less than $T_{\rm disso}$).
 We shall study the some properties of this system at $T=0$ and 
compare them with those of a general dilute
Bose-Fermi mixture.
In particular, we shall be able to find the effective interaction, or
equivalently the scattering length, between the 
Bosons and Fermions in terms of $a$.
We further show that the Bogoliubov mode of the system
is the same as that of a Bose-Fermi mixture.

In the case where the pairing between the spin-up and down
particles are in the Cooper limit 
(weak attractive interaction so that $a < 0$ and small),  
the system with $n_{\uparrow} > n_{\downarrow}$ has been studied
before in the context of superconductivity under an applied
Zeeman field $h$ \cite{Sarma}.  We shall compare 
our results with those of this better known system.

We now consider the system at $T = 0 $ under the mean-field
approximation
 The necessary 
mean field equations
can be obtained by generalizing those of ref \cite{Leggett80}
(which deals with $n_{\uparrow} = n_{\downarrow}$).
We introduce three fields, the chemical potentials 
$\mu_{\uparrow}$, $\mu_{\downarrow}$ for the two spin species
and a pairing field $\Delta$.  These fields obey the
self-consistent equations

\be
1 = { 4 \pi \hbar^2 a \over m}
\int_{\vec k} \ 
\left[    { 1 \over 2 \epsilon_k }   -
  { 1 \over 2 E_k } \ \left( 1 - f(E_k-h) - f(E_k + h) \right)
\right]
\label{delta}
\ee

\be
n_{\sigma} = \int_{\vec k} \ n_{\sigma}(\vec k) \qquad ,
\qquad  \sigma = {\uparrow} \ {\rm or} \ {\downarrow}
\label{ntot}
\ee 

\noindent with

\be
n_{\uparrow}(\vec k) = 
u_k^2 \  f(E_k -h) \ + \ v_k^2 \  ( 1 - f(E_k + h) )
\label{nup}
\ee

\be
n_{\downarrow}(\vec k) =
u_k^2 \ f(E_k + h) \ + \ v_k^2 \ ( 1 - f(E_k - h) )
\label{ndown}
\ee

\noindent where $f$ is the Fermi function.
We have defined for convenience 
the chemical potential average and difference 
$\mu \equiv  (\mu_{\uparrow} + \mu_{\downarrow})/2$,
$h \equiv  (\mu_{\uparrow} - \mu_{\downarrow}) / 2 $ ($>0$),
the energies
$\epsilon_k \equiv \hbar^2 k^2/2m$,
 $E_k \equiv [ (\epsilon_k - \mu)^2 + |\Delta|^2 ] ^{1/2}$,
and the "coherence factors" 
$u_k^2 = {1 \over 2} ( 1 + {\epsilon_k - \mu \over E_k} )$,
$v_k^2 = {1 \over 2} ( 1 - {\epsilon_k - \mu \over E_k} )$,
$u_k v_k = {\Delta \over 2 E_k } $
are exactly as in BCS theory for superconductivity.
Eq (\ref{delta}) is the self-consistent equation for $\Delta$
whereas eqs (\ref{ntot}) are the 
constraints for the Fermion concentrations,
with $n_{\sigma}(\vec k)$ the occupation number at wavevector
$\vec k$.
Notice that if we view the system as a Bose-Fermi mixture,
the chemical potential for the Bosons is $ \mu_b = 2 \mu$ 
and that for the (left-over) Fermions is $\mu_{\uparrow}$. 
In the dilute limit $0 < x_{\uparrow} \equiv n_{\uparrow} a^3 << 1$
and  $0 < x_{\downarrow} \equiv n_{\downarrow} a^3 << 1$
which we are studying (in contrast to the Cooper limit where $a< 0$ ),
it is convenient to study the solution to eqs (\ref{delta})-(\ref{ntot})
as an expansion in $x_{\uparrow}$ and $x_{\downarrow}$.
At $T=0$, $f(E_k + h) = 0$ and $f(E_k - h)$ is non-vanishing
only for $k < k_c$ where $E_{kc} = h$.  
From eq (\ref{nup}) and (\ref{ndown}) we get immediately

\be
\nu = k_c^3 / 6 \pi^2 \ .
\label{nu}
\ee

\noindent $k_c$ plays the role of 
the Fermi wavevector for the excess spin-up Fermions
(which form a `Fermi sea', see also below)

We shall see (as in the corresponding case of ref \cite{Leggett80})
that $ \mu < 0$ and $|\Delta| <<  |\mu|$.  Using this inequality,
eq(\ref{delta}) and eq(\ref{ndown}), we obtain
two equations determinating $\Delta$ and $\mu$.
In the dilute limit we get 

\be
 \mu = - {\hbar^2 \over 2 m a^2} + { 2 \pi ( n_b + 2 \nu) \hbar^2 a \over m}
\label{mu}
\ee

\be
|\Delta|^2 = 8 \pi n_b \epsilon_b^2 a^3  \ 
[ 1  -  ... ]
\label{delta2}
\ee

\noindent (and hence $|\Delta|^2 / \mu^2 \sim n_b a^3 << 1$ as 
promised).
Eq (\ref{mu}) can be interpreted easily
by recalling $\mu_b = 2 \mu$.
The first term arises from the binding energy of
the molecule.  The two parts of the 
second term represent the mean-field
corrections to the chemical potential due to
the interaction of the molecules among themselves and
with the Fermions respectively.
Equating these two expressions with
($1/2$ times) $n_b g_{bb}$ and $\nu g_{bf}$, 
we obtain the interaction constants 

\ba
g_{bb} &=& { 4 \pi \hbar^2 a \over m}  
\label{gbb} \\
g_{bf} &=& { 8 \pi \hbar^2 a \over m}
\label{gbf}
\ea

\noindent or equivalently the scattering lengths 
$a_{bb} = 2 a$ and $a_{bf} = 8a / 3$. 
\cite{appendix}

\begin{figure}[h]
\epsfig{figure=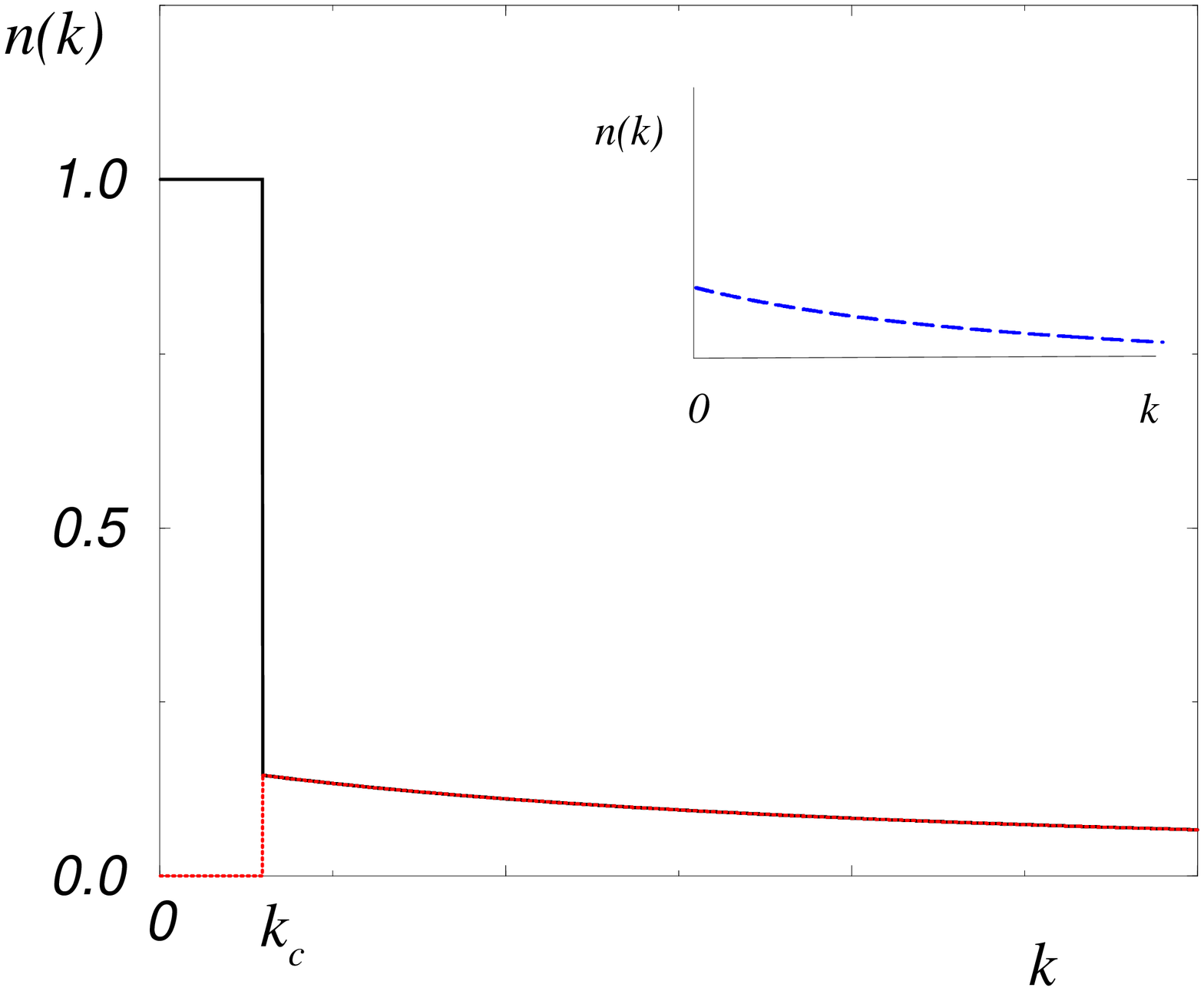,width=3.5in,angle=-90}
\vskip 0.4 cm
\caption[]{ Schemetic plot of $n_{\sigma}(\vec k)$ 
versus $k$.  Full line: $n_{\uparrow}(\vec k)$,
dotted line: $n_{\downarrow}(\vec k)$.
The magnitude of $n_{\sigma}(\vec k)$ for
$k > k_c$ has been exaggerated for clarity.  Inset:
Sketch of $n(\vec k)$ for the case of $n_{\uparrow} = n_{\downarrow}$.}

\label{fig:nk}
\end{figure}

We now discuss some properties associated
with the Fermions. 
The occupation numbers of the constituent Fermions 
$n_{\sigma}(\vec k)$ are sketched in Fig. \ref{fig:nk} for $T=0$.
$n_{\uparrow} (k)  = 1$ whereas $n_{\downarrow} (k)  = 0$ for $k < k_c$.   
$n_{\uparrow} (k) $ and $n_{\downarrow} (k) $ 
are equal but small for $k > k_c$.  These results
can be understood by considering the wavefunction of the sytem
\cite{wf}.  

It is also interesting to study the spectral function
$A_{\uparrow}(\vec k, \omega)$ 
($=  - { 1 \over \pi} \ {\rm Im} \ G^R _{\uparrow 11} (\vec k, \omega)$
given below)
 corresponding to the density
of states for addition/removal of a spin-up Fermion at momentum $\vec k$
and energy $\omega$.  $A_{\uparrow} (\vec k, \omega)$ has
the same formal expression as in the case of Cooper pairing
(c.f. eq (\ref{G11}) below):
$A (\vec k, \omega) = u_k^2 \delta ( \omega - (E_k - h))
  + v_k^2 \delta ( \omega + (E_k + h))$. There are
thus two delta-function peaks, one at positive and one at
negative energies. Since 
$|\Delta| << \mu$, $v_k^2 << 1$ and thus the weight
at the negative frequency is very small.  To a 
very good approximation, $A_{\uparrow} \sim \delta (\omega - (\epsilon_k
- \mu_{\uparrow}))$, that of a free Fermi gas.
We shall compare these results with those of case (B) below.

We next study the dynamic properties of our system (at $T=0$) 
by evaluating the two particle vertex function (matrix)
${\bf \Gamma} (Q, \Omega)$ for repeated scattering between
 a spin-up and spin-down
particle (hole) at small wavevector $Q$ and frequency $\Omega$.
  We shall see that,
as in the case of $n_{\uparrow} = n_{\downarrow}$, 
\cite{Melo93,Haussmann99}
${\bf \Gamma}$ is related to the propagator for the Bosons.
 ${\bf \Gamma}$, in Matsubara frequencies (we shall
drop the superscript $M$ for simplicity.  The desired
${\bf \Gamma}$ in real frequency can be obtained via
$i \Omega_{\nu} \to \Omega $ with $\Omega_{\nu} > 0$ as usual)
 obeys the Bethe-Salpeter equation
(generalizing that of  \cite{Haussmann99})

\be
\Gamma^{-1} _{\alpha1, \alpha2} (Q, \Omega_{\nu}) = 
  { m \over 4 \pi \hbar^2 a} \delta_{\alpha1, \alpha2}
+ M _{\alpha1, \alpha2} (Q, \Omega_{\nu})
\label{BS}
\ee
where
\be
M _{\alpha1, \alpha2} (Q, \Omega_{\nu})
\equiv
\int_{\vec k}  \  \left[ \  T \sum_{\epsilon_n} \  
G_{\uparrow \ \alpha1, \alpha2} ( \vec Q - \vec k, \Omega_{\nu} - \epsilon_n)
G_{\downarrow \ \alpha1, \alpha2} (  \vec k, \epsilon_n)
- { m \over \hbar^2 k^2 } \delta_{\alpha1, \alpha2} \ 
\right]  
\label{M}
\ee

\noindent and $G_{\sigma \  \alpha1, \alpha2}$
are the matrix Green's function in particle ($\alpha = 1$)
and hole ($\alpha = 2$) space for spin-$\sigma$ 
 defined in a similar manner as in Gorkov's 
theory of superconductivity \cite{Fetter},
 and $\Omega_{\nu}$ and $\epsilon_n$
are Bosonic and Fermionic Matsubara frequencies respectively.
In principle there is a contribution to the self-energy for ${\bf G}$ proportional
to ${\bf \Gamma}$ and therefore
 ${\bf G}$ and ${\bf \Gamma}$ have to be found self-consistently 
(see \cite{Haussmann99} ).
However, at $T=0$ and under our diluteness condition
we can simply use the mean-field solution for $G$,
{\it i.e.},

\be
G_{\uparrow 11} (\vec k, \epsilon_n) = 
{u_k^2 \over i \epsilon_n - ( E_k - h) }   +
{v_k^2 \over i \epsilon_n + ( E_k + h) }
\label{G11}
\ee
and
\be
G_{\uparrow 12} (\vec k, \epsilon_n) = 
u_k v_k \left[ {1 \over i \epsilon_n - ( E_k - h) }   -
{ 1 \over i \epsilon_n + ( E_k + h) } \right]
\ee
where $u_k$, $v_k$ are the coherence factors introduced before.
$G_{\downarrow}$'s are given by similar expressions except
$h \to - h$ and $\Delta \to - \Delta$.
${\bf \Gamma}^{-1} (Q=0, \Omega_{\nu}=0)$ can be calculated easily and 
 is of order $\Delta^2$ due to
eq (\ref{delta}).  However, evaluation of ${\bf \Gamma}^{-1}$ at
finite $Q$ and $\Omega$ needs special care.
In $M_{11}$ there arises the contribution
\ba
M^{a}_{11} (Q, \Omega)
&\equiv& - \int_{\vec k} \ 
u^2_{\vec Q - \vec k} v^2_{\vec k} \ 
T \sum_{\epsilon_n}
\left[
 {1 \over i \epsilon_n + ( E_{\vec Q - \vec k} - h) - \Omega }   \times
{ 1 \over i \epsilon_n + ( E_{\vec k} - h) } \right]
\nonumber \\
& = & 
\int_{\vec k} \
u^2_{\vec Q - \vec k} v^2_{\vec k} \
{ [ 1 - sgn(( E_{\vec Q - \vec k} - h) - \Omega ) 
       sgn(E_{\vec k} - h ) ] / 2
  \over 
 | E_{\vec Q - \vec k} - E_{\vec k } - \Omega | }  \qquad .
\nonumber         
\ea
The integrand is singular as $Q$ and $\Omega$ approaches $0$.
To evalute $M^{a}_{11}$, we first note that
under our diluteness approximation, we can replace
$E_k $ by $\epsilon_k - \mu$ and thus
$E_k - h  = E_k - E_{kc} \approx \epsilon_k - \epsilon_{kc}$ etc.
The integrand is therefore finite only if
$\epsilon_{\vec Q - \vec k_c} - \epsilon_{k_c} - \Omega$ and
$\epsilon_k - \epsilon_{k_c}$  are of opposite signs, and thus 
for $ Q << k_c$, $\Omega << k_c^2/m$, 
 $\vec k$ is restricted to be near the "Fermi surface",
i.e., $|\vec k| \sim k_c$.  We can then use
$u^2_{\vec Q - \vec k} v^2_{\vec k} \approx u^2_{kc} v^2_{kc}
\approx \left( { |\Delta| \over \epsilon_b } \right)^2$,
and replace the $\vec k$
integral by integration over $\hat k$ and 
$\xi_k \equiv \epsilon_k - \epsilon_{k_c}$.  
We finally obtain 
$M^a_{11} (Q, \Omega) \approx { m k_c \over 2 \pi^2}
  ({|\Delta| \over \epsilon_b})^2 $, and hence

\be
- \Gamma^{-1}_{11} (Q, \Omega) = 
 { 1 \over 8 \pi \epsilon_b^2 a^3 } \ 
\left\{
- \Omega + { Q^2 \over 4 m}
  + {|\Delta|^2 \over 2 \epsilon_b}
 -  8 \pi |\Delta|^2 a^3  
\left( { m k_c \over 2 \pi^2 } \right) 
\right\}  \ \ .
\label{Gamma11}
\ee
\noindent Similarly
\be
- \Gamma^{-1}_{12} (Q, \Omega) = 
 { 1 \over 8 \pi \epsilon_b^2 a^3 } \ 
\left\{
   {\Delta^2 \over 2 \epsilon_b}
 -  8 \pi \Delta^2 a^3 
\left( { m k_c \over 2 \pi^2 } \right) 
\right\} \ \ .
\label{Gamma12}
\ee

As in the case of equal number of spin-up and down particles,\cite{Haussmann99}
${\bf \Gamma}$ is thus found to be proportional to the 
connected part of the Bosonic propagator matrix ${\bf G}_b$.  
${\bf G}_b^{-1}$, under the random phase approximation,
reads (c.f. \cite{Yip01})
\be
- G_{b 11}^{-1} (Q, \Omega) 
= - \Omega + { Q^2 \over 2 m_b}
   + g_{bb} n_b
   - g^2_{bf} n_b \chi_f(Q,\Omega)
\label{GB11}
\ee
\be
- G_{b 12}^{-1} (Q, \Omega) 
=  + g_{bb} n_b
   - g^2_{bf} n_b \chi_f(Q,\Omega)
\label{GB12}
\ee
where $\chi_f$ is the density response of the Fermions and
we have taken the gauge where the phase of the condensate is zero.
To further check our claim that 
${\bf \Gamma} \propto {\bf G}_b$ above,
we note that $m_b = 2 m$,
$g_{bb} n_b = {|\Delta|^2 \over 2 \epsilon_b} $,
$g_{bf}^2 n_b =  8 \pi |\Delta|^2 a^3 $
(using eqs (\ref{delta2}), (\ref{gbb}) and (\ref{gbf})), 
$\chi_f (Q, \Omega) = \left( { m k_c \over 2 \pi^2 } \right) $.
In this last expression, we have used the value of 
$\chi_f (Q, \Omega)$ for $ \Omega / Q \to 0$.  That this is 
the correct limit can be seen as follows.
The bare sound velocity for the Bosons is given by
$c_b^0 = ( n_b g_{bb} / m_b ) ^{1/2}$ whereas the
bare Fermi velocity is $ v_f^0 = k_c / m$.  Thus
$ (c_b^0 / v_f^0 )^2  \sim ( n_b a / k_c^2 ) << 1$ 
under our diluteness condition.  For the
mode under consideration, $ (\Omega / v_f Q ) \sim (c_b^0 / v_f^0 ) << 1$.

In conclusion, we have studied, via many-body technique, some properties
of an atom-molecule mixture where the Fermionic atoms are
identical with one of the constituent atoms
of the diatomic molecules. 
If  $T << T_{BEC}$,
our system behaves like a general Bose-Fermi mixture.
In this case 
we have also obtained the scattering lengths
$a_{bb}$ and $a_{bf}$ for our mixture in terms of $a$.

I would like to thank Darwin Chang for pointing out and 
making available to me ref \cite{Haussmann99}.
This research was supported by the National Science
Council of Taiwan under grant number 89-2112-M-001-061.

\begin{thebibliography}{12}
\bibitem{wf}  The (unrenormalized) wavefunction for our system can be written
as (generalizing \cite{Leggett80})
$ \left( \Pi_{\vec k', k' < k_c} c^{\dagger}_{\vec k' \uparrow} \right) \times
\left( \sum_{\vec k} \chi(k) c^{\dagger}_{\vec k \uparrow}
c^{\dagger}_{-\vec k \downarrow} \right)^{N_b}   |{\rm vac} > $.
Here $c^{\dagger}_{\vec k \sigma}$ is the creation operator of
a Fermion with wavevector $\vec k$ and spin $\sigma$,
$\chi(k) = u_k v_k $ the (Fourier transformed) spatial part of the
molecular wavefunction, and $N_b$ the total number of Bosons.
\end{thebibliography}
\end{document}